\documentclass[sigconf]{acmart}
\def\BibTeX{{\rm B\kern-.05em{\sc i\kern-.025em b}\kern-.08emT\kern-.1667em\lower.7ex\hbox{E}\kern-.125emX}}
    
\copyrightyear{2019}
\acmYear{2019}
\setcopyright{none}
\acmConference[]
\acmBooktitle{}
\usepackage[utf8]{inputenc} % allow utf-8 input
\usepackage[T1]{fontenc}    % use 8-bit T1 fonts
\usepackage{hyperref}       % hyperlinks
\usepackage{url}            % simple URL typesetting
\usepackage{booktabs}       % professional-quality tables
\usepackage{amsfonts}       % blackboard math symbols
\usepackage{nicefrac}       % compact symbols for 1/2, etc.
\usepackage{microtype}      % microtypography
\usepackage{graphicx}
\usepackage{amsmath}

\begin{document}

\title{Adversarial Attacks Against Medical Deep Learning Systems}

%
% The "author" command and its associated commands are used to define the authors and their affiliations.
% Of note is the shared affiliation of the first two authors, and the "authornote" and "authornotemark" commands
% used to denote shared contribution to the research.

\author{Samuel G. Finlayson}
\email{samuel\_finlayson@hms.harvard.edu}
\affiliation{%
  \institution{Department of Systems Biology and MD-PhD program \\ Harvard Medical School and MIT}
  %\streetaddress{10 Shattuck Street}
  \city{Boston}
  \state{MA}
  %\postcode{02115}
}

\author{Hyung Won Chung}
\email{hwc@mit.edu}
\affiliation{%
  \institution{Massachusetts Institute of Technology}
  %\streetaddress{1 Th{\o}rv{\"a}ld Circle}
  \city{Cambridge}
  \state{MA}
  %\country{Iceland}
  }

\author{Isaac S. Kohane}
\email{isaac\_kohane@hms.harvard.edu}
\affiliation{%
  \institution{Department of Biomedical Informatics \\ Harvard Medical School}
  %\streetaddress{10 Shattuck Street}
  \city{Boston}
  \state{MA}
  %\postcode{02115}
}

\author{Andrew L. Beam}
\email{andrew\_beam@hms.harvard.edu}
\affiliation{%
  \institution{Department of Biomedical Informatics \\ Harvard Medical School}
  %\streetaddress{10 Shattuck Street}
  \city{Boston}
  \state{MA}
  %\postcode{02115}
}

%
% By default, the full list of authors will be used in the page headers. Often, this list is too long, and will overlap
% other information printed in the page headers. This command allows the author to define a more concise list
% of authors' names for this purpose.
\renewcommand{\shortauthors}{Finlayson et al.}

%
% The abstract is a short summary of the work to be presented in the article.
\begin{abstract}
The discovery of adversarial examples has raised concerns about the practical deployment of deep learning systems. In this paper, we demonstrate that adversarial examples are capable of manipulating deep learning systems across three clinical domains. For each of our representative medical deep learning classifiers, both white and black box attacks were highly successful. Our models are representative of the current state of the art in medical computer vision and, in some cases, directly reflect architectures already seeing deployment in real world clinical settings. In addition to the technical contribution of our paper, we synthesize a large body of knowledge about the healthcare system to argue that medicine may be uniquely susceptible to adversarial attacks, both in terms of monetary incentives and technical vulnerability. To this end, we outline the healthcare economy and the incentives it creates for fraud and provide concrete examples of how and why such attacks could be realistically carried out. We urge practitioners to be aware of current vulnerabilities when deploying deep learning systems in clinical settings, and encourage the machine learning community to further investigate the domain-specific characteristics of medical learning systems.
\end{abstract}

% NOTE: DONE BY ANDY


%
% Keywords. The author(s) should pick words that accurately describe the work being
% presented. Separate the keywords with commas.

%
% A "teaser" image appears between the author and affiliation information and the body 
% of the document, and typically spans the page. 
%  \includegraphics[width=\textwidth]{make_an_aa.png}
%  \Description{Enjoying the baseball game from the third-base seats. Ichiro Suzuki preparing to bat.}

%
% This command processes the author and affiliation and title information and builds
% the first part of the formatted document.
\maketitle

\section{Introduction}
Over the past seven years, deep learning has transformed computer vision and has been implemented in scores of consumer-facing products. Many are excited that these approaches will continue to expand in scope and that new tools and products will be improved through the use of deep learning. One particularly exciting application area of deep learning has been in clinical medicine. There are many recent high-profile examples of deep learning achieving parity with human physicians on tasks in radiology \cite{gale2017detecting,rajpurkar2017chexnet}, pathology \cite{bejnordi2017diagnostic}, dermatology \cite{esteva2017dermatologist}, and opthalmology \cite{gulshan2016development}. In some instances, the performance of these algorithms exceed the capabilities of most individual physicians in head-to-head comparisons. This has lead some to speculate that entire specialties in medical imaging, such as radiology and pathology, may be radically reshaped \cite{jha2016adapting} or cease to exist entirely. Furthermore, on April 11, 2018, an important step was taken towards this future: the U.S. Food and Drug Administration announced the approval of the first computer vision algorithm that can be utilized for medical diagnosis without the input of a human clinician \citep{PressAnn97:online}.

In parallel to this progress in medical deep learning, the discovery of so-called `adversarial examples' has exposed vulnerabilities in even state-of-the-art learning systems  \citep{goodfellow2014explaining}. Adversarial examples -- inputs engineered to cause misclassification -- have quickly become one of the most popular areas of research in the machine learning community \citep{Szegedy2013, nguyen2015deep, moosavi2016deepfool, papernot2016distillation}. While much of the interest with adversarial examples has stemmed from their ability to shed light on possible limitations of current deep learning methods, adversarial examples have also received attention because of the cybersecurity threats they may pose for deploying these algorithms in both virtual and physical settings \citep{papernot2016distillation, melis2017deep, kurakin2016adversarial, athalye2017synthesizing, brown2017adversarial, grosse2017adversarial}. 

Given the enormous costs of healthcare in the US, it may seem prudent to take the expensive human `out of the loop' and replace him or her with an extremely cheap and highly accurate deep learning algorithm. This seems especially tempting given a recent study's finding that physician and nursing pay is one of the key drivers of high costs in the US relative to other developed countries \cite{papanicolas2018health}. However, there is an under-appreciated downside to widespread automation of medical imaging tasks given the current vulnerabilities of these algorithms. If we seriously consider taking the human doctor completely `out of the loop' (which now has legal sanction in at least one setting via the FDA, with many more to likely follow), we are forced to also consider how adversarial attacks may present new opportunities for fraud and harm. In fact, even with a human in the loop, any clinical system that leverages a machine learning algorithm for diagnosis, decision-making, or reimbursement could be manipulated with adversarial examples.

In this paper, we extend previous results on adversarial examples to three medical deep learning systems modeled after the state of the art medical classifiers. On the basis of these results and knowledge of the healthcare system, we argue that medical imaging is particularly vulnerable to adversarial attacks and that there are potentially enormous incentives to motivate prospective bad actors to carry out these attacks. We hope that by highlighting these vulnerabilities, more researchers will explore potential defenses against these attacks within the healthcare domain. Because the healthcare system is complex and administrative processes can appear byzantine, it may be difficult to imagine how these attacks could be operationalized. To ground these abstract potentials for harm in actual use cases, we describe realistic scenarios where clinical tasks might rely on deep learning and give specific examples of the fraud that could be mediated by adversarial attacks.  Our goal is to provide background on the distinct features of the medical pipeline that make adversarial attacks a threat, and also to demonstrate the practical feasibility of these attacks on real medical deep learning systems. 

\subsection{Adversarial Examples}

Adversarial examples are inputs to machine learning models that have been crafted to force the model to make a classification error. This problem extends back in time at least as far as the spam filter, where systematic modifications to email such as `good word attacks' or spelling modifications have long been employed to try to bypass the filters \citep{lowd2005adversarial, lowd2005good, dalvi2004adversarial}. More recently, adversarial examples were re-discovered and described in the context of deep computer vision systems through the work of \citet{Szegedy2013} and \citet{goodfellow2014explaining}. Particularly intriguing in these early examples was the fact that adversarial examples could be crafted to be extremely effective despite being imperceptibly different from natural images to human eyes. In the years since, adversarial examples -- visible and otherwise -- have been shown to exist for a wide variety of classic and modern learning systems \citep{papernot2016transferability, brown2017adversarial, elsayed2018adversarial}. By the same token, adversarial examples have been extended to various other domains such as text and speech processing \citep{jia2017adversarial, carlini2016hidden}. For an interesting history of adversarial examples and methods used to combat them see \citet{biggio2017wild}.

\begin{figure*}[!htb]
  \centering
    \includegraphics[width=\textwidth]{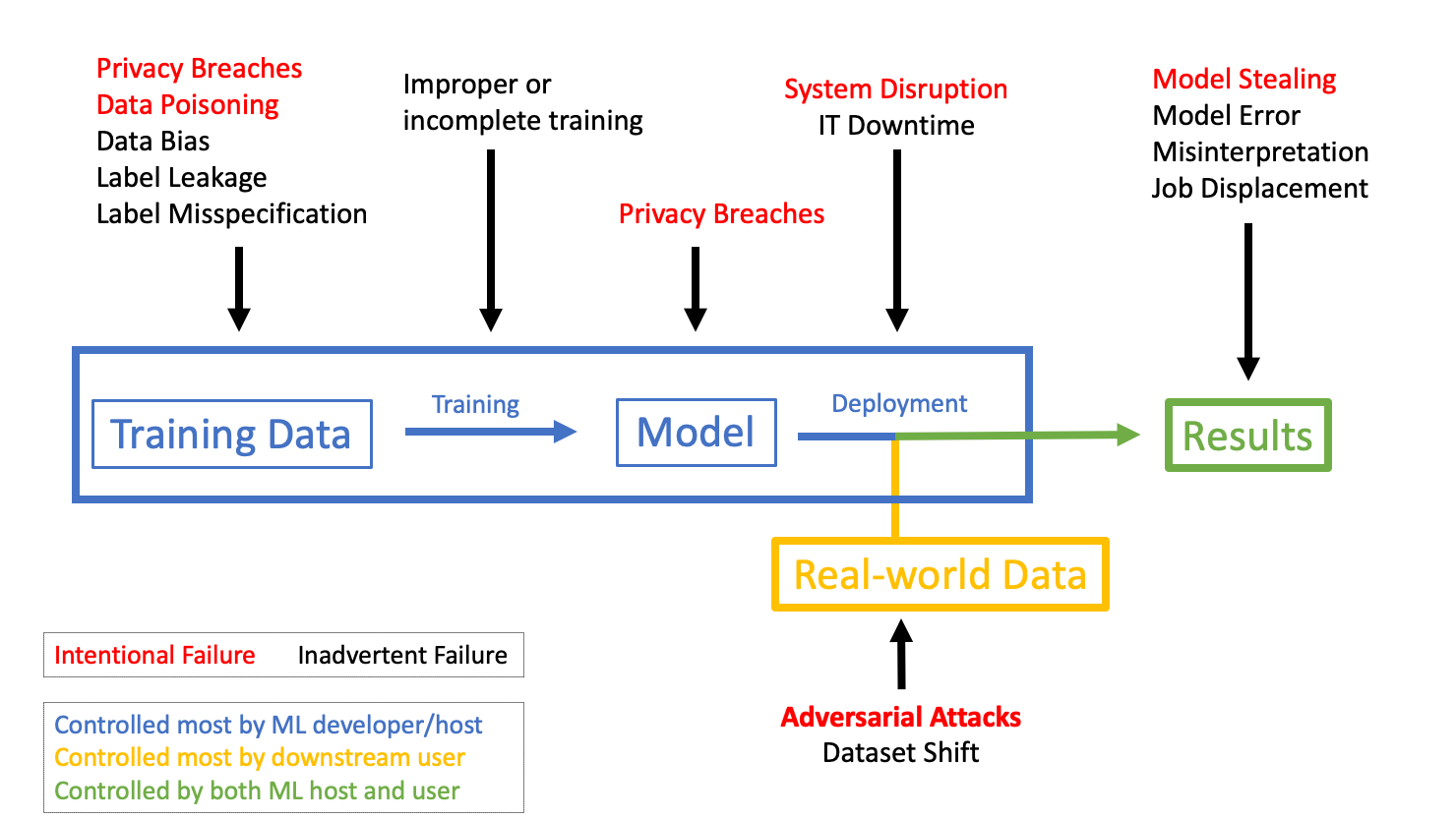}
    \caption{\label{fig:attack_taxonomy} Adversarial attacks within the broader taxonomy of risks facing the machine learning pipeline. Adversarial attacks pose just one of many possible risks in the development and deployment of ML systems, but are noteworthy because they enable end-users to manipulate model outputs without ever influencing the training process or gaining access to the deployed model itself.}
\end{figure*}

Figure ~\ref{fig:attack_taxonomy} places adversarial attacks within the broader context of machine learning development and deployment. While there are many possible intentional and inadvertent failures of real-world machine learning systems, adversarial attacks are particularly important to consider from the standpoint of model deployment, because they enable those submitting data into a running ML algorithm to subtly influence its behavior without ever achieving direct access to the model itself or the IT infrastructure that hosts it.

Adversarial examples are generally thought to arise from the piecewise linear components of complex nonlinear models \citep{goodfellow2014explaining}. They are not random, they are not due to overfitting or incomplete model training, they occupy only a comparatively small subspace of the feature landscape, they are robust to random noise, and they have been shown to transfer in many cases from one model to another \cite{papernot2016transferability, Tramer2017}. Furthermore, in addition to executing many successful attacks in purely virtual settings, researchers in the past several years have also demonstrated that adversarial attacks can generalize to physical world settings \citep{kurakin2016adversarial, evtimov2017robust, athalye2017synthesizing, brown2017adversarial}.

One natural question raised by the existence of adversarial examples is to what extent and in what forms they constitute a viable risk for harm in real-world machine learning settings. Many authors have discussed the feasibility of and possible motivations for adversarial attacks on certain real-world systems such as self-driving cars \citep{evtimov2017robust, carlini2017towards, lu2017no}. However, to our knowledge, previous machine learning literature has yet to thoroughly discuss the possibility of adversarial attacks on medical systems.

\section{Identifying factors in the U.S. healthcare system that favor adversarial attacks}

In this section, we provide a synthesis of aspects of the healthcare system that may create both the incentive and the opportunity for a bad actor to carry out an adversarial attack.

\subsection{Background on the healthcare economy and possible incentives for fraud via adversarial attacks}

\textbf{The healthcare economy is huge and fraud is already pervasive}. The United States spent approximately \$3.3 trillion (17.8\% of GDP) on healthcare in 2016 \cite{papanicolas2018health}, and healthcare is projected to represent 1/5 of the US economy by 2025. Given the vast sums of money represented by the healthcare economy, inevitably some seek to profit by committing fraud against the healthcare system. Medical fraud is estimated to cost hundreds of billions of dollars each year, and one study estimated this number to be as high as \$272 billion in 2011 \cite{jain2014corruption}. Fraud is committed both by large institutions and by individual actors. Large institutions engage in fraud by systemically inflating costs for services to increase revenue \cite{jama_fraud_def, rudman2009healthcare}. Likewise, it has been found that some individual physicians routinely bill for the highest allowable amount over 90\% of the time \cite{charles_ornstein_2018}. 

\textbf{Algorithms will likely make medical reimbursement decisions in the future.}  Due to the amount of money involved with the delivery of healthcare, complex book-keeping systems have been created to facilitate billing and reimbursement. In fact, most of the data generated by the healthcare system in the electronic healthcare record (EHR) is created to justify payments from `payers' (private or public insurers) to 'providers' (hospitals and physicians). In many cases, the level of monetary reimbursement for a given patient hinges on the establishment of specific diagnostic `codes', which are used to record a patient's diagnoses and treatments with high granularity \footnote{See for instance the International Classification of Disease code V97.33XD, which represents the following diagnosis: `Sucked into jet engine, subsequent encounter'}\citep{sanders2012road}. In an effort to increase revenue, some providers engage in the practice of 'upcoding' diagnoses or procedures -- selecting the codes which will allow them to bill for the highest amount. For their part, insurance companies seek to minimize total expenditure by investing millions of dollars in IT and personnel to identify unjustified billing codes. The resultant struggle between payors and providers has been extensively documented \citep{kesselheim2005overbilling, wynia2000physician}. To ensure consistency and justifiability, insurance companies will often demand specific gold standard tests as proof of diagnosis before reimbursing a given medical claim, and leverage increasingly sophisticated analytics to determine reimbursement value. Given these dynamics, it is seemingly inevitable that insurance companies will begin to require algorithmic confirmation of certain diagnoses before providing reimbursement. If and when this occurs, the ability to undetectably influence (either as a provider or as a payer) the outputs of trusted and otherwise unbiased diagnostic systems would result in the ability to influence the movement of billions of dollars through the healthcare economy. Even today, the practice of upcoding, which often entails finding subtle combinations of codes that influence reimbursement \citep{reynolds2005metabolic}, can itself be arguably considered a form of adversarial attack against reimbursement algorithms.

\textbf{Algorithms will increasingly determine pharmaceutical and device approvals.} The monetary value of a successful clinical trial is immense, with one recent study estimating the median revenue across individual cancer drugs to be as high as \$1.67 billion only four years after approval \citep{prasad2017research}. At the same time, regulatory bodies such as the FDA are increasingly allowing for the approval of new drugs based on digital \textit{surrogates} for patient response, including medical imaging  \citep{pien2005using}. As algorithmic endpoints for clinical trials become increasingly accepted -- and deep learning algorithms continue to assert themselves as equal or superior to humans on well-defined visual diagnostic tasks -- we could soon reach a future where billion dollar drug decisions are made primarily by machines. In such a future, effectively executed adversarial attacks could allow trialists to imperceptibly `put their thumb on the scale,' even if images are vetted to ensure they are coming from the correct patients.

\subsection{Distinctive technical sources of vulnerability to adversarial attacks among medical machine learning systems}

\textbf{Ground truth is often ambiguous.} Compared to most common image classification tasks, the ground truth in medical imaging is often controversial, with even specialty radiologists disagreeing on well defined tasks \citep{njeh2008tumor, li2009variability, brouwer20123d}. As such, if end-users selectively perturb images for which it is difficult to establish the true diagnosis, they can make it extremely difficult to detect their influence through even expert human review. Somewhat ironically, it is these borderline cases are where deep learning is likely to be most valuable.

\textbf{Medical imaging is highly standardized.} Compared to other domains of computer vision, medical imaging is extremely standardized, with images generally captured with pre-defined and well-establisehd positioning and exposure \citep{simon1971principles}. As such, medical adversarial attacks do not need to meet the same standards of invariance to lighting and positional changes as attacks on other real-world systems such as self-driving cars. This is potentially important, as some have argued that dynamic viewing conditions imply that `there is a good prospect that adversarial examples can be reduced to a curiosity with little practical impact' \cite{lu2017no}.

\textbf{Commodity network architectures are often used.} Nearly all of the most successful published methods in medical computer vision have consisted of the same fundamental architecture: one of a small set of pretained ImageNet models that was fine-tuned to the specific task \cite{gulshan2016development,wang2017chestx, esteva2017dermatologist}. This lack of architectural diversity could make it easier for potential attackers to build transferable attacks against medical systems. By the same token, given the importance of peer review and publication in validating and approving medical diagnostics, it is likely the architectures of most medical image classification models will be public for the sake of transparency, allowing for more targeted adversarial attacks.

\textbf{Medical data interchange is limited and balkanized}. Five electronic health record (EHR) vendors constitute about half of the market and hundreds of others serve the other half. Even within an EHR vendor, data sharing is spotty and the terminologies and their semantics vary considerably from one implementation to another. On the one hand, this means that there are no universally shared  mechanisms for authentication, verification of message integrity, data quality metrics, nor mechanisms for automated oversight. On the other hand, this allows healthcare providers to customize their EHR's, billing, and other information technology systems in ways that are opaque to most external auditors using one-size-fits-all tools and methods.

\textbf{Hospital infrastructure is very hard to update.} Medical software is often implemented within monolithic enterprise-wide proprietery software systems, making updates, revisions and fixes expensive and time-consuming. For context, consider the coding dictionaries used to classify patients' diseases, the International Classification of Disease (ICD) system. As recently as 2013, most hospitals were operating using the ninth edition of this coding scheme, published in 1978, despite the fact that a revised version (ICD-10) was published in 1990. All told, the conversion to the ICD-10 coding scheme has been estimated to cost major health centers up to \$20 million \textit{each} and require up to 15 years \citep{sanders2012road}. Others decided in the early 2000s that it would be preferable to skip the 1990 schema entirely and wait for ICD-11, despite the fact that its release wasn't scheduled until 2018. Thus, vulnerabilities present in medical software are likely to persist for years due to the difficulty and expense involved with update hospital infrastructure. 

% For further context in this regard, one can consider that the primary means of transferring text records between hospitals continues to be the \textit{fax machine}. When it comes to images, a patient trying to complete the simple task of obtaining images to send to a new physician will in many cases have to write a letter and wait weeks for a manual image export to a disk drive.  This is not an industry that could roll out complex image pipeline updates overnight.

\textbf{Medicine contains a mix of technical and non-technical workers.} Compared to many other industries, medicine is extremely interdisciplinary and mostly comprised of members who lack a strong computational or statistical training background. For example, in the case of self-driving cars, the teams developing computer vision systems are likely to be led and staffed primarily by engineers, if not computer scientists. In contrast, since the clinical usability of medical imaging systems is also extremely important, hospitals are likely to lean heavily on physician-researchers in developing these systems, who tend to lack robust computational expertise \cite{manrai2014medicine}.

\textbf{Biomedical images carry personal signatures that could be used to defend against many simpler attacks, but not against adversarial examples.} One potential alternative to adversarial attacks on part of malicious end-users would be to simply substitute in true images of the target class. However, biomedical images -- including retinal images, X-rays, and skin photographs -- are often as unique to their owners as fingerprints \citep{qamber2012personal, angyal1998personal, miller2010personal}. This provides several advantages against substitution attacks: First, algorithms could be designed to check input images against prior images from the same patient, to ensure that their identities match. Second, algorithms could query against the database of previously submitted images and flag any inputs that are likely from the same patients as previous entries. Both strategies would make it more difficult for fraudsters to continually execute substitution attacks against ML systems, but neither would defend against adversarial attacks, which needn't change the personal identifiers in the image. By analogy, adversarial attacks applied to billing codes or medical text could serve to manipulate reimbursement algorithms by selecting combinations of codes or words that are individually truthful but in combination yield anomalously high reimbursement.

\textbf{There are many potential adversaries.} The medical imaging pipeline has many potential attackers and is thus vulnerable at many different stages.  While in theory one could devise elaborate image verification schemes throughout the data processing pipeline to try to guard against attacks, this would be extremely costly and difficult to implement in practice.

\section{Attacking representative clinical deep learning systems}

To demonstrate the feasibility of medical adversarial attacks, we developed a series of medical classifiers modeled after state-of-the-art clinical deep learning systems, and launched successful white- and black-box attacks against each. 

\subsection{Methods}
\subsubsection{Construction of medical classification models}

 We developed baseline models to classify referable diabetic retinopathy from retinal fundoscopy (similar to \citet{gulshan2016development}), pneumothorax from chest-xray (similar to \citet{wang2017chestx} and \citet{rajpurkar2017chexnet}), and melanoma from dermoscopic photographs (similar to \citet{esteva2017dermatologist}). The decision to build models for these particular tasks was made both due to public data availability, as well as the fact that they represent three of the most highly visible successes for medical deep learning.

All of our models were trained on publicly available data. For diabetic retinopathy, this was the Kaggle Diabetic Retinopathy dataset \citep{kaggle}. The key distinction with the Kaggle dataset, however, was that we were seeking to predict \textit{referable} (grade 2 or worse) diabetic retinopathy in accordance with \citet{gulshan2016development} rather than predicting the retinopathy grade itself as was the case in the Kaggle competition. As such, the training and test sets from the kaggle competition were merged, relabeled using their provided grades, and split by patient into training and test sets with probability 0.88/0.12. For the chest x-rays, we used the ChestX-Ray14 dataset described by \citet{wang2017chestx}. We identified cases and controls by selecting images whose labeled contained `pneumothorax' or `no finding', respectively, and additionally excluded from our control group any images from patients who had received both labels. We then split by patient into training and test sets with probability 0.85/0.15. For melanoma, we downloaded images labeled as benign or malignant melanocytic lesions from the International Skin Imaging Collaboration website \citep{isic}, splitting again into training and test sets with probability 0.85/0.15.

As in the case of all three of the original papers that inspired these models, we built our classifiers by fine-tuning a pretrained ImageNet model. For convenience and consistency, we chose to build each of our networks using a pretrained ResNet-50 model, fine-tuned in Keras using stochastic gradient descent with a learning rate of 1E-3 and Momentum of 0.9. Data was augmented using 45$^{\circ}$ rotation and horizontal flipping chest x-ray images, and with 360$^{\circ}$ rotation, vertical and horizontal flipping, and mixup \cite{zhang2017mixup} for fundoscopy and dermoscopy images. In all three cases, these settings provided respectable performance and we therefore didn't perform dedicated hyperparameter optimization.

% \begin{table}[t]
%   \label{results-table1}
%   \centering
%   \begin{tabular}{lllll}               \\
%     \midrule
%     \textbf{Inputs}   & \textbf{Accuracy}  &  \textbf{AUC}  &  \textbf{Sensitivity}  & \textbf{Specificity} \\
%     \midrule
%     Fundoscopy  & 91\% & 0.91 & 85\% & 80\%\\
%     Chest X-Ray & 95\% & 0.94 & 90\% & 82\%\\
%     Dermoscopy & 88\% & 0.86 & 80\% & 73\%\\
%     \midrule
%   \end{tabular}
%   \caption{Results of baseline medical AI models on the validation set.}
% \end{table}

\subsubsection{Construction of adversarial attacks}

To demonstrate the vulnerability of our models to adversarial attacks under a variety of threat models, we implemented both human-imperceptible and patch attacks.

For our human-imperceptible attacks, we followed the white and black box PGD attack strategies published in \citet{madry2017towards}, which were also described as baselines in \citet{kannan2018adversarial}. The PGD attack (see \citet{madry2017towards} and also \citet{kurakin2016adversarial}) is an iterative extension of the canonical fast gradient sign method (FGSM) attack developed by \citet{goodfellow2014explaining}. In the PGD attack, given input $x\in \mathbb{R}^d$, loss function $L(\theta, x, y)$, and a set of allowed perturbations $\mathcal{S} \subseteq \mathbb{R}^d$ (most commonly the $\ell_\infty$ ball around $x$), one can perform a projected gradient descent on the negative loss function:
\begin{align*} 
x^{t+1} = \Pi_{x+\mathcal{S}} (x^t + \epsilon \textrm{sgn}(\nabla_x L(\theta, x, y))) 
\end{align*}
in order to identify an optimal perturbation. We implemented the PGD attack using the library Cleverhans, conducting 20 iterations with hyperparameter $\epsilon=0.02$ ($\epsilon$ corresponds to the maximum permitted $\ell_\infty$ norm of the perturbation) \citep{papernot2016cleverhans}.

For our adversarial patch attacks, we followed the approach of \citet{brown2017adversarial}. The learning process of the adversarial patch $\hat{p}$ uses a variant of the expectation over transformation algorithm (originally proposed by \citet{Athalye2017}), which can be expressed as the following maximization:

\begin{equation}
    \hat{p} = \underset {p}{\operatorname {arg\,max}}\ \mathbb{E} \left[ \log p_{Y|X}\left( \left. \hat{y} \right| A\left(p, X, L, T \right) \right) \right]
\end{equation}

where $p_{Y|X}(\cdot|\cdot)$ represents the probability output from the classifier given the input image $X$, $L$ is the location of the patch, $T$ is the transformation (rotation and scaling), $\hat{y}$ is the target label (for our application binary label), and $A(\cdot, \cdot, \cdot, \cdot)$ is the deterministic mapping of given image, patch, location and transformation into the adversarially patched input image. The expectation is over the locations, transformations and input images, thus allowing the resulting patch to be `universal,' because it was trained over the entire training set and robust since training averaged over various location and transformations. We deployed adversarial patches with scaling parameter 0.4. 

White box attacks for each attack were implemented by building attacks directly on the victim model itself. As in \citet{madry2017towards} and \citet{kannan2018adversarial}, black box attacks were performed by crafting the attack against an independently-trained model with the same architecture and then transferring the resultant adversarial examples to the victim.
% While this form of black box attack is slightly `gray' due to the fact that we knowingly duplicate the overall architecture of the model, the attack makes no use of any weights or predictions generated by the victim model itself. Further, given the standard of publishing and publicly vetting medical systems prior to approval, it is entirely sensible to assume that many or most attackers will have knowledge of the architecture of the systems they attack.

Finally, as a control, we implemented a naive patch attack using natural images, as in \citet{brown2017adversarial}. To make this as strong a baseline as reasonably possible, we built our natural image patches using the images assigned by the model with the highest probability of the target class. Resultant patches were then applied to the test set images with random rotation and scaling factor 0.4, so match the adversarial patch.

\textbf{Availability of code:} Code to reproduce the analyses and results can be found at the first author's Github account: \\
\texttt{https://github.com/sgfin/adversarial-medicine}.

\subsection{Results}

\begin{figure*}[!htb]
  \centering
    \includegraphics[width=\textwidth]{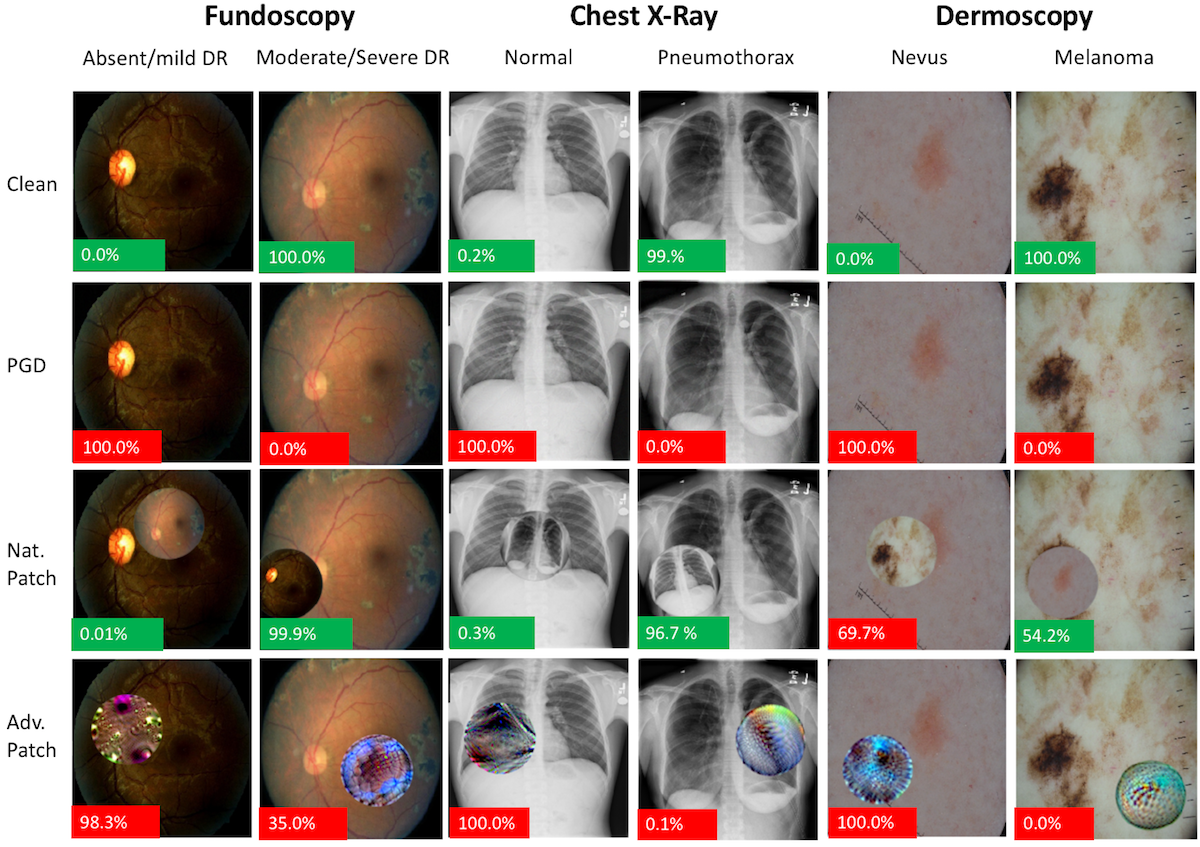}
    \caption{\label{fig:attack_results} Characteristic results of adversarial manipulation. Each clean image represents the natural image to which the model assigns the highest probability for the given diagnosis. The percentage displayed on the bottom left of each image represents the probability that the model assigns that image of being diseased. Green = Model is correct on that image. Red = Model is incorrect.}
\end{figure*}

The results of our experiments are depicted in Table~\ref{results-table} and in Figure ~\ref{fig:attack_results}.

While discrepancies in data source and train-test partitioning makes direct comparison to state-of-the-art models unfeasible, all of our baseline models achieved performance reasonably consistent with the results reported in the original manuscripts on natural images: AUROC of 0.910 for diabetic retinopathy compared with 0.936 reported in \citet{gulshan2016development}, AUROC of 0.936 for pneumothorax compared with 0.90 reported by \citet{rajpurkar2017chexnet}, and AUROC of 0.86 on melanoma compared with 0.91-0.94 reported in \citet{esteva2017dermatologist}.

Projected gradient descent attacks, targeting the incorrect answer in every case, produced effective AUROCs of 0.000 and accuracies of 0\% for all white box attacks. Black box attacks produced  AUROCs of less than 0.10 for all tasks, and accuracies ranging from 0.01\% on fundoscopy to 37.9\% on dermoscopy. Qualitatively, all attacks were human-imperceptible.

Adversarial patch attacks also achieved effective AUROCs of 0.000 and accuracies of <1\% for white box attacks on all tasks. Black box adversarial patch attacks achieved AUROCs of less than 0.005 for all tasks and accurracies less than 10\%.  The "natural patch" controls created by adding patches created from the most-strongly classified image of the desired class resulted in AUROCs ranging from 0.48-0.83 with accuracies ranging from 67.5\% to 92.1\%.

\begin{table*}[ht]
\centering
\scalebox{1.0}{%
\begin{tabular}{l|ccc|ccc|ccc}
\hline
 & \multicolumn{3}{c|}{\textit{Fundoscopy}} & \multicolumn{3}{c|}{\textit{Chest X-Ray}} & \multicolumn{3}{c|}{\textit{Dermoscopy}} \\
\multicolumn{1}{c|}{Input Images} & \multicolumn{1}{l}{Accuracy} & \multicolumn{1}{l}{AUROC} & \multicolumn{1}{l|}{Avg. Conf.} & \multicolumn{1}{l}{Accuracy} & \multicolumn{1}{l}{AUROC} & \multicolumn{1}{l|}{Avg. Conf.} & \multicolumn{1}{l}{Accuracy} & \multicolumn{1}{l}{AUROC} & \multicolumn{1}{l|}{Avg. Conf.} \\
\midrule
Clean & 91.0\% & 0.910 & 90.4\% & 94.9\% & 0.937 & 96.1\% & 87.6\% & 0.858 & 94.1\% \\
PGD - White Box & 0.00\% & 0.000 & 100.0\% & 0.00\% & 0.000 & 100.0\% & 0.00\% & 0.000 & 100.0\% \\
PGD - Black Box & 0.01\% & 0.002 & 90.9\% & 15.1\% & 0.014 & 92.6\% & 37.9\% & 0.071 & 92.0\% \\
Patch - Natural &  78.5\% & 0.828 & 80.8\%  &  92.1\% & 0.539 & 95.8\% & 67.5\% & 0.482 & 85.6\% \\
Patch - White Box &  0.3\% & 0.000 & 99.2\%  &  0.00\% & 0.000 & 98.8\%  &  0.00\% & 0.000 & 99.7\%  \\
Patch - Black Box &  3.9\% & 0.000 & 97.5\%  &  9.7\% & 0.004 & 83.3\%  &  1.37\% & 0.000 & 97.6\% \\
\bottomrule
\end{tabular}%
}
\caption{Results of medical deep learning models on clean test set data, white box, and black box attacks.}
\label{results-table}
\end{table*}

\section{Discussion}

Our experiments indicate that adversarial attacks are likely to be feasible even for extremely accurate medical classifiers, regardless of whether prospective attackers have direct access to the model or require their attacks to be human imperceptible. Of note, while the PGD attacks require digital access to the specific images to be sent into the model, adversarial patch attacks are universal in the sense that they can be applied to any image. This could open the possibility for implementation of attacks upstream from the image capture itself, rendering data processing defenses such as image hashing at point-of-capture ineffective. In addition, it is noteworthy to recognize that adversarial patches were far more potent than "photoshop"-style natural patch attacks that alter the image using the most strongly classified image from the training set.

We now discuss how someone might perform adversarial attacks against the systems developed in previous section under a realistic set of conditions. As stated above, we focus this discussion on imaging-based ML models, though similar arguments apply to adversarial examples crafted on billing code submissions or medical text. For the purposes of illustration, consider a scenario where the ML models have been subjected to extensive testing and validation and are now clinically deployed. These systems would function much like laboratory tests do now and provide confirmation of suspected diagnoses. In some instances, an insurance company may require a confirmatory diagnosis from one of these systems in order for a reimbursement to be made. Further, the insurance company or regulatory agency may have separate methods deployed to ensure the patient identity matches from prior images or has never been submitted from the same provider before. We provide the examples below to show that in many instances there is both the \emph{opportunity} and \emph{incentive} for someone to use an adversarial example to defraud the healthcare system. 

\subsection{Hypothetical examples}
\textbf{Adversarial examples in dermatology}: Dermatology in the US operates under a `fee for service' model wherein a physician or practice is paid for the services or procedures they perform for the patient. Under this model, dermatologists are incentivized to perform as many procedures as possible, as their revenue is directly tied to the amount of procedures they perform. This has caused some dermatologists to perform a huge number of unnecessary procedures to increase revenue. For example, one dermatologist in Florida was recently sentenced to 22 years in prison after performing more than three thousand unnecessary surgical procedures \cite{rudman2009healthcare}. To combat fraud and unnecessary procedures such as this, an insurance company could require that a deep learning system (e.g. the one from Section 4) analyzes all dermoscopy images to confirm that surgery is necessary. In this scenario, a bad actor could add adversarial noise to images to ensure that the deep learning model always gives the diagnosis that he or she desires. Furthermore, they could add this noise to `borderline' cases, which would render the attack nearly impossible to detect by human review. Thus, a bad actor like the Florida dermatologist from \cite{rudman2009healthcare} could sidestep an insurance company's image-based fraud detector and continue to defraud the system in perpetuity.

\textbf{Adversarial examples in radiology}. Thoracic radiology images (typically CT scans, which is a 3D application of X-Ray technology) are also often used to measure tumor burden, a common secondary endpoint of cancer therapy response\citep{pien2005using}. To foster more rapid and more universally standardized clinical trials, the FDA might consider requiring that trial endpoints, such as tumor burden in chest imaging, be evaluated by a deep learning system such as the one from Section 4. By applying undetectable adversarial perturbations to the images, a company running a trial could effectively guarantee a positive trial result with respect to this endpoint, even if images are subsequently released to the public for inspection. In addition, chest X-rays provide a common screening test for dozens of diseases, and a positive chest X-ray result is often used to justify more heavily reimbursed procedures such as biopsies, CT or MR imaging, or surgical resection. As such, one could imagine many scenarios arising around chest X-rays that are directly analogous to the melanoma detection situation described above.

\textbf{Adversarial examples in ophthalmology}. As described in Section 3, providers and pharmaceutical companies are not the only organizations that could be incentivized to employ adversarial manipulation. Often, the entities who pay for healthcare (such as private or public insurers) wish to curtail the utilization rates of certain procedures to reduce costs. However, there are often guidelines from government agencies (such as the Centers for Medicare and Medicaid Services) that specify diagnostic criteria which if present dictate that certain procedures must be covered. One such criterion could be that any patient with a confirmed diabetic retinopathy diagnosis from a deep learning system such as the one from Section 4 must have the resulting vitrectomy surgery covered by their insurer. Even though the insurer has no ability to control the policy, they could still control the rate of surgeries by applying adversarial noise to mildly positive images, reducing the number of procedures. On the other end of the spectrum, an ophthalmologist could affix a universal adversarial patch to the lens of his image capture system, forcing a third party image processing system to mistake all images for positive cases without having to make an alterations to the image within the IT system itself.

\subsection{Possible areas for further research}
We hope that our discussion and demonstrations can help motivate further research into adversarial examples generally as well as within the specific context of healthcare. In particular, we consider the following areas to be of high priority: 

\textbf{Algorithmic defenses} against adversarial examples remain an extremely open and challenging problem. We defer a full discussion of the extremely rapidly evolving field of adversarial defenses to the primary literature \citep{kannan2018adversarial, goodfellow2014explaining, kolter2017provable, madry2017towards, papernot2016transferability, Szegedy2013, Tramer2017, papernot2016distillation, kolter2017provable, dvijotham2018dual, raghunathan2018certified, buckman2018thermometer}. Unfortunately, despite the explosive emergence of defense strategies, there does not appear to be a simple and general algorithmic fix for the adversarial problem available in the short term. For example, one recent analysis investigated a series of promising methods that relied on gradient obfuscation, and demonstrated that they could be quickly broken \citep{athalye2018obfuscated}. Despite this, we also note that principled approaches to adversarial robustness are beginning to show promise. For example, several papers have demonstrated what appears to be both high accuracy and strong adversarial robustness on smaller datasets such as MNIST,  \citep{madry2017towards,kannan2018adversarial}, and there have also been several results including theoretical \textit{guarantees} of adversarial robustness, albeit on small datasets and/or with still-insufficient accuracy \citep{kolter2017provable}. Generalized attempts at algorithmic robustness are promising, but have yet to provide methods that can demonstrate high levels of both accuracy and adversarial robustness at ImageNet scale. However, \textbf{domain-specific algorithmic} defenses such as dataset-specific image preprocessing have been shown to be highly effective on some datasets \citep{graese2016assessing}. In this light, particularly given the highly standardized image capture procedures in biomedical imaging, we feel that medical-domain-specific algorithmic defenses offer an important and promising area of future research.

\textbf{Infrastructural defenses} against clinical adversarial attacks include methods deployed to prevent potential bad actors from altering medical images -- or at least make it easier to confirm image tampering if adversarial examples are suspected. For example, imaging devices could immediately store a hashed version of any image they generate, which could subsequently be used as a reference. Likewise, raw clinical images could be processed and analyzed on a third-party system to prevent any possible systemic manipulation by payers or providers. This family of approaches to standardized best practices is reminiscent of the system of Clinical Laboratory Improvement Amendments (CLIA), a set of federal policies that regulates the process by which clinical laboratory samples are handled and analyzed in the United States \citep{us1992medicare}. Given that algorithmic defenses against adversarial attacks are still very much an area of research, we feel that infrastructural defenses should be strongly considered for all medical classifier systems that could carry incentives for adversarial attacks. However, implementing healthcare system-wide standardization to this end represents and immense challenge that will require buy-in from both the medical and CS communities.

\textbf{Ethical tradeoffs} are introduced by adversarial examples. As outlined above, several papers have demonstrated greater improvements in adversarial robustness that come at a cost of lower accuracy \citep{paschali2018generalizability, tsipras2018robustness}. However, in medical imaging, this introduces an ethical conundrum: how does one weigh protecting against adversarial examples against any inaccurate diagnosis? Quantifying and making this trade-off explicit would allow for more informed decision making and system design.

\section{Conclusion}

The prospect of improving healthcare and medicine with the use of deep learning is truly exciting. There is reasonable cause for optimism that these technologies can improve outcomes and reduce costs, if judiciously implemented \cite{beam2016translating}. In this light, it is unsurprising that dozens of private companies and large health centers have initiated efforts to deploy deep learning classifiers in clinical practice settings. As such efforts continue to develop, it seems inevitable that medical deep learning algorithms will become entrenched in the already multi-billion dollar medical information technology industry. However, the massive scale of the healthcare economy brings with it significant opportunity and incentive for fraudulent behavior and ultimately, patient harm. 

In this work, we have outlined the systemic and technological reasons that cause adversarial examples to pose a disproportionately large threat in the medical domain, and provided examples of how such attacks may be executed. We hope that our results help facilitate a discussion on the threat of adversarial examples among both computer scientists and medical professionals. For machine learning researchers, we recommend research into infrastructural and algorithmic solutions designed to guarantee that attacks are infeasible or at least can be retroactively identified. For medical providers, payers, and policy makers, we hope that these practical examples can motivate a meaningful discussion into how precisely these algorithms should be incorporated into the clinical ecosystem despite their current vulnerability to such attacks.

\begin{acks}
The authors would like to thank Aleksander Madry, Dimitris Tsipras, Yun Liu, Jasper Snoek, Alex Wiltschko for a helpful review of our manuscript. In addition, SGF was supported by training grants T32GM007753 and T15LM007092; the content is solely the responsibility of the authors and does not necessarily represent the official views of the National Institute of General Medical Sciences or the National Institutes of Health.
\end{acks}

%
% The next two lines define the bibliography style to be used, and the bibliography file.
\bibliographystyle{ACM-Reference-Format}
\bibliography{main}

% 
% If your work has an appendix, this is the place to put it.
% \appendix
% \section{Additional adversarial examples}

% \begin{figure*}[!htb]
%   \centering
%     \includegraphics[width=\textwidth]{appen_dr.png}
%     \caption{\label{fig:appen_dr} Additional examples of retinal fundus images, adversarial perturbations, and resultant adversarial examples. Noise displayed is the white box noise.}
% \end{figure*}

% \begin{figure*}[!htb]
%   \centering
%     \includegraphics[width=\textwidth]{appen_cx.png}
%     \caption{\label{fig:appen_cx} Additional examples of chest x-ray images, adversarial perturbations, and resultant adversarial examples. Noise displayed is the white box noise.}
% \end{figure*}

% \begin{figure*}[!htb]
%   \centering
%     \includegraphics[width=\textwidth]{appen_derm.png}
%     \caption{\label{fig:appen_derm} Additional examples of dermoscopy images, adversarial perturbations, and resultant adversarial examples. Noise displayed is the white box noise.}
% \end{figure*}

\end{document}